\documentclass[iop]{emulateapj}
\usepackage{epsfig}
\usepackage{csquotes}
\usepackage{graphicx}
\usepackage{ulem,color}
\usepackage[dvipsnames]{xcolor/xcolor}
\usepackage{dcolumn}
\usepackage{bm}
\usepackage{longtable}
\usepackage{natbib}
\usepackage[bookmarks=true]{hyperref}
\usepackage{xspace}
\usepackage{amsmath}
\def\lesssim{\mathrel{\hbox{\rlap{\hbox{\lower4pt\hbox{$\sim$}}}\hbox{$<$}}}}
\def\gtrsim{\mathrel{\hbox{\rlap{\hbox{\lower4pt\hbox{$\sim$}}}\hbox{$>$}}}}

\newcommand{\bea}{\begin{eqnarray}}
\newcommand{\eea}{\end{eqnarray}}

\newcommand{\bP}{{\bf P}}
\newcommand{\bF}{{\bf F}}
\newcommand{\bU}{{\bf U}}

\newcommand{\prim}{{{\mathbf{P}}}}

\newcommand{\harm}{{\sc Harm3d}\xspace}

\def\lambdabar{%
\relax
\bgroup
\def\@tempa{\hbox{\raise.73\ht0
\hbox to0pt{\kern.25\wd0\vrule width.5\wd0
height.1pt depth.1pt\hss}\box0}}%
\mathchoice{\setbox0\hbox{$\displaystyle\lambda$}\@tempa}%
{\setbox0\hbox{$\textstyle\lambda$}\@tempa}%
{\setbox0\hbox{$\scriptstyle\lambda$}\@tempa}%
{\setbox0\hbox{$\scriptscriptstyle\lambda$}\@tempa}%
\egroup
}

\slugcomment{Submitted to ApJ}

\begin{document}

\title{Quasi-Periodicity of Supermassive Binary Black Hole Accretion Approaching Merger}

\author{Dennis B. Bowen    $^{1,2}$,
  Vassilios Mewes $^2$,
  Scott C. Noble $^{3,4}$,
  Mark Avara $^{2}$,
  Manuela Campanelli $^2$, and
  Julian H. Krolik $^5$
} 

\affil{$^1$ Center for Theoretical Astrophysics and X Computational Physics,\\ Los Alamos National Laboratory,
  Los Alamos, New Mexico 87545\\
  $^2$ Center for Computational Relativity and Gravitation,
  Rochester Institute of Technology, Rochester, NY 14623\\
  $^3$ Department of Physics and Engineering Physics,
  The University of Tulsa, Tulsa, OK 74104\\
  $^4$ NASA Postdoctoral Program Senior Fellow, Goddard Space Flight Center, Greenbelt, MD 20771\\
  $^5$ Department of Physics and Astronomy, Johns Hopkins
  University, Baltimore, MD 21218\\
}

\email{dbowen@lanl.gov}

\begin{abstract}
In this paper we continue the first ever study of magnetized mini-disks
coupled to circumbinary accretion in a supermassive binary black hole (SMBBH)
approaching merger reported in~\cite{Bowen18}. We extend this simulation from
3 to 12 binary orbital periods. We find that relativistic SMBBH
accretion acts as a resonant cavity, where quasi-periodic oscillations
tied to the the frequency at which the black hole's orbital phase matches
a non-linear $m=1$ density feature, or ``lump'', in the circumbinary accretion
disk permeate the system.
The rate of mass accretion onto each of the mini-disks around the black holes is modulated
at the beat frequency between the binary frequency and the lump's mean orbital frequency, i.e.,
$\Omega_{\rm beat} = \Omega_{\rm bin} - \bar{\Omega}_{\rm lump}$, while the total mass accretion rate
of this equal-mass binary is modulated at two different frequencies, $\gtrsim \bar{\Omega}_{\rm lump}$ and
$\approx 2 \Omega_{\rm beat}$.  The instantaneous rotation rate of the lump
itself is also modulated at two frequencies close to the modulation frequencies of the
total accretion rate, $\bar{\Omega}_{\rm lump}$ and $2 \Omega_{\rm beat}$.
Because of the compact nature of the mini-disks in SMBBHs approaching merger,
the inflow times within the mini-disks are comparable to the period on which their mass-supply varies,
so that their masses---and the accretion rates they supply to their black holes---are
strongly modulated at the same frequency.
In essence, the azimuthal symmetry of the circumbinary disk is broken by the dynamics of
orbits near a binary, and this $m=1$ asymmetry then drives quasi-periodic variation
throughout the system, including both accretion and disk-feeding.  In SMBBHs approaching
merger, such time variability could introduce distinctive, increasingly rapid, fluctuations
in their electromagnetic emission.

\end{abstract}


\section{Introduction}
\label{sec:introduction}
In addition to the stellar-mass binary black holes (BBHs) already detected
by LIGO~\citep{GWdetect,Abbott:2016nmj,PhysRevLett.118.221101,2017ApJ...851L..35A,LIGOScientific:2018mvr,LIGOScientific:2018jsj},
supermassive binary black holes (SMBBHs) are
expected to form during galactic mergers (see\citep{Khan16,Kelley17} for more details).
While there remains uncertainty about the exact processes which 
extract sufficient angular momentum from the binary to create separations
of $\mathcal{O}\left(10^3\right)$ gravitational radii~\citep{BBR80},
interactions with both stars~\citep{Khan11,Vasiliev15,2017MNRAS.464.2301G} and
gas~\citep{Shi12,2012AdAst2012E...3D} may overcome
this so-called ``final parsec problem'' (but see, e.g., \citet{Munoz2019,Moody2019}). Beyond this point,
gravitational wave emission can be expected to drive the binary to merger
within a Hubble time\citep{MP05}. Unfortunately,
the detection of gravitational waves from SMBBHs awaits the efforts of
other observational campaigns, such as LISA~\citep{2017arXiv170200786A} 
and Pulsar Timing
Arrays~\citep{SHANNON15}. 

However, the galactic environments in which SMBBHs form
should provide ample amounts of gas for them to be electromagnetically bright\citep{Cuadra09,Chapon13,Colpi14}.
With this consideration, and the plethora of current and upcoming electromagnetic
survey campaigns~\citep{2009arXiv0912.0201L,2016arXiv161205560C}, SMBBHs represent an excellent candidate
for multimessenger astrophysics following the launch of LISA~\citep{2019arXiv190305287K,2019arXiv190304417B}.
However, the chief question remains: ``What are the electromagnetic signals associated with SMBBH merger?''.
To zeroth order, any electromagnetic counterpart to the SMBBH gravitational wave signal will be
directly related to the amount and structure of gas in the immediate vicinity of the black holes
(BHs) to be gravitationally heated at merger~\citep{Krolik10}.

The earliest estimates predicted that SMBBHs would exhibit a dry merger, in which
there would be little to no gas available for electromagnetic emission.
When the binary mass ratio is near unity, torques exerted by the binary on inflowing gas open up
a cavity of radius $\approx 2a$, where $a$ is the binary separation~\citep{P91,ArtymLubow94,ArtymLubow96}.
It was thought that this process also prevented any mass from crossing the cavity.
However, more recent numerical simulations have shown that in fact the BHs efficiently peel
streams of gas off the inner edge of circumbinary disks, {\it and these streams rapidly traverse}
the central cavity\citep{MM08,2010ApJ...715.1117B,Pal10,Farris11,Bode12, Farris12,Giacomazzo12,Noble12,Shi12,DOrazio13,Farris14,Gold14,Farris15,Farris15a,Shi2015,DOrazio16,Bowen18,2018MNRAS.476.2249T}.

Moreover, these streams of gas are able to form
individual accretion disks, or mini-disks, around
each member of the binary\citep{Farris14,Farris15,Farris15a,DOrazio16,2018MNRAS.476.2249T,Moody2019}.
Because the mini-disks represent some of the most significant departures from standard
active galactic nuclei, much effort has been undertaken to understand
their structure~\citep{Ju16,Bowen17,RyanMacFadyen17,Bowen18} and mechanisms by which
they could produce distinct electromagnetic signatures~\citep{Roedig:2014,2015Natur.525..351D,Bowen17,RyanMacFadyen17,Bowen18,dAscoli:2018fjw,2018MNRAS.476.2249T}.

Finally, numerical studies have shown that the inner edge of the circumbinary disk itself differs
significantly from a standard accretion disk around a single
BH~\citep{MM08,Noble12,Shi12,DOrazio13,Farris14,Farris15,Farris15a,DOrazio16,2018MNRAS.476.2249T}.
As streams of gas are peeled off the inner edge of the circumbinary disk, a portion of the stream is granted
angular momentum from the binary and flung into the inner edge of the circumbinary disk.
This serves as an initial seed for a feedback cycle resulting in an $m=1$ azimuthal
asymmetry, or lump, at the inner edge of the circumbinary disk~\citep{Shi12}.
The lump quasi-periodically modifies the accretion flux into
the central cavity and therefore onto the mini-disks\citep{Shi12,Farris14,Farris15,Farris15a,DOrazio16,Bowen18}.

This paper picks up from where the equal-mass binary simulation of
\cite{Bowen18} left off, and seeks to explore the time-dependent
structure of relativistic mini-disks for binary separations of
$a \lesssim 20M$\footnote{We adopt geometrized units with $G=c=1$
where distance has units of $GM/c^2$ and $M$ is the total mass of the binary.}.
The lump is located $\approx 2.4a$ from the center-of-mass and orbits at nearly the
local Keplerian orbital frequency around the center-of-mass, ${\bar\Omega}_{\rm lump} \approx 0.28 \Omega_{\rm bin}$,
where $\Omega_{\rm bin}$ is the binary orbital frequency. The quasi-periodic modulations
of the accretion into the central cavity due to the lump occur
at twice the frequency of a BH coming into phase with the lump:
$2\Omega_{\rm beat} = 2\left(\Omega_{\rm bin} - \bar{\Omega}_{\rm lump}\right)$
\citep{Shi12,Noble12,DOrazio13,Farris14,Farris15,Farris15a,DOrazio16,Bowen18}. 

While Newtonian studies do note the modulation of the accretion streams, the inflow times in
the mini-disks are sufficiently long that no significant mini-disk asymmetry was achieved~\citep{Farris14,Farris15,Farris15a,DOrazio16}.
However, once the binary separation shrinks to $a \lesssim 20M$, the tidal truncation radius
of a mini-disk is only a factor of a few times the innermost stable circular orbit (ISCO). This scenario
leads to radial pressure gradients accelerating mass inflow 
well beyond the rate associated with stresses arising from MHD turbulence \citep{BI2001,KHH05} and in turn
dynamically couples the mini-disk mass to the lump~\citep{Bowen18}.

By extending the original simulation, we now numerically extract the modulation frequencies of the mini-disk
masses, combined mini-disk mass, and lump orbital frequency. 
We find that nearly every component of
the lump-mini-disk system exhibits quasi-periodic modulations associated with the
frequency at which the BH's orbital phase matches the lump's:
$\Omega_{\rm beat} \approx 0.72\Omega_{\rm bin}$.
For instance, we observe that the mini-disk masses have a modulation frequency of $\Omega_{\rm beat}$.
Meanwhile, both the combined mini-disk mass and the instantaneous lump angular velocity $\Omega_{\rm lump}$
experience modulation at frequencies
$\approx \left( 0.2 - 0.4\right) \Omega_{\rm bin}$ and $2\Omega_{\rm beat}$.
We therefore conclude
that the internal time-dependent structure of the central cavity, streams, lump, and mini-disks are all
intimately coupled together. Such rich time variability could introduce significant quasi-periodic signatures
in the electromagnetic emission of SMBBHs~\citep{dAscoli:2018fjw}.

The remainder of this paper is organized as follows.
In Section~\ref{sec:simulation-details} we present the details of our simulation. In Section~\ref{sec:results}
we present the results of our analysis including the extraction of characteristic frequencies of oscillations
in the lump-mini-disk system. In Section~\ref{sec:discussion} we discuss the implications of our
findings. Finally, in Section~\ref{sec:conclusions} we present our concluding remarks.

\section{Simulation Details}
\label{sec:simulation-details}
\subsection{Overview}
The primary objective of our simulation is to further explore the
dynamic coupling of the mini-disks around each BH to a
$m=1$ azimuthal Fourier mode in the circumbinary disk reported in~\citep{Bowen18}.
We continue the simulation from three binary orbital periods, where
\cite{Bowen18} ended, out to 12 binary orbital periods.
Our simulation starts at a binary separation of $20M$, where relativistic
contributions to the spacetime and inspiral are significant~\citep{2015PhRvD..91b4034Z}, and the
circumbinary disk is taken from a snapshot of $t=50,000M$ in runSE of~\cite{Noble12}.

As the total mass in the accretion disks for astrophysical
SMBBHs will be negligible compared to the combined mass of the binary, 
we neglect the self-gravity of the fluid and any feedback into the spacetime from the matter. We
approximate the spacetime of the SMBBH with a fully analytic
prescription which asymptotically matches BH perturbation theory
near each individual BH to post-Newtonian (PN) theory at
2.5PN order (for full details see~\cite{Noble12,Proj0,Ireland:2015cjj,Bowen17}).
The BH trajectories are calculated to 3.5PN~order accuracy.
We evolve the gas using the equations of general relativistic magnetohydrodynamics (GRMHD)
in flux-conservative form using the \harm code (see~\cite{Noble09,Noble12,Bowen17}).

Throughout the paper, unless otherwise noted, we use
geometrized units in which $G=c=1$.  When used as tensorial indices,
we reserve Greek letters (e.g., $\alpha, \beta, \gamma, \ldots$) for
spacetime indices and Roman letters (e.g., $i, j, k, \ldots$) as
indices spanning spatial dimensions.

\subsection{General Relativistic Magnetohydrodynamics}
\label{sec:EOM}
The equations of motion for GRMHD on a background spacetime are expressed
through conservation laws for baryon number density and stress-energy coupled
to the Maxwell induction equations and the divergence free constraint
on the magnetic field. These may be written as the following system of conservation laws:
\begin{equation}
\partial_t \bU\left(\prim\right) =                                                                       
-\partial_i \bF^i\left(\prim\right) + \mathbf{S}\left(\prim\right) \ ,
\label{cons-form-mhd}
\end{equation}
where $\bP$ are the ``primitive'' variables, $\bU$ the ``conserved''
variables, $\bF^i$ the fluxes, and $\mathbf{S}$ the source terms. In
terms of the primitive variables and metric functions they can be
expressed as
\begin{eqnarray}
\bU\left(\prim\right) & = & \sqrt{-g} \left[ \rho u^t ,\, {T^t}_t +
  \rho u^t ,\, {T^t}_j ,\, B^k\right]^T \ , \label{cons-U-mhd} \\
\bF^i\left(\prim\right) & = & \sqrt{-g} \left[ \rho u^i ,\, {T^i}_t +
  \rho u^i ,\, {T^i}_j ,\, \left(b^i u^k - b^k u^i \right)\right]^T \ , \label{cons-flux-mhd} \\
\mathbf{S}\left(\prim\right) & = & \sqrt{-g} \left[ 0 ,\,
  {T^\kappa}_\lambda {\Gamma^\lambda}_{t \kappa} - \mathcal{F}_t ,\,
  {T^\kappa}_\lambda {\Gamma^\lambda}_{j \kappa} - \mathcal{F}_j ,\, 0
  \right]^T , \label{cons-source-mhd}
\end{eqnarray}    
where $g$ is the determinant of the metric, ${\Gamma^\lambda}_{\alpha \beta}$ 
are the Christoffel symbols,  $b^{\alpha} = \left(1/u^t\right)\left({\delta^\alpha}_\nu + u^\alpha u_\nu \right)B^\nu$ is the magnetic 
4-vector projected into the fluid's comoving reference frame, and $u^{\alpha}$ are the components of the fluid's
4-velocity. The stress-energy tensor is written as
\begin{equation}
  T_{\alpha \beta} = \left( \rho h + 2p_m \right) u_{\alpha} u_{\beta} + \left( p + p_m\right)g_{\alpha \beta} - b_{\alpha}b_{\beta} \ ,
\end{equation}
where $h = 1 + \epsilon + p/\rho$ is the specific enthalpy, $\epsilon$ is the specific
internal energy, $p$ is the gas pressure, 
$p_m = \frac{1}{2}b^2$ is the magnetic pressure, 
and $\rho$ is the rest-mass density. The
initial value of the divergence constraint is maintained to machine
precision using FluxCT~\citep{2000JCoPH.161..605T}.

We assume the accretion disks are radiatively efficient and cool
away any local increases in entropy on a timescale $t_{\mathrm cool}$,
specified as the local fluid orbital period (for full details on
how these are calculated see~\citep{Bowen17,dAscoli:2018fjw}). We include cooling
as a source term to the stress-energy conservation equations: $\nabla_\lambda {T^{\lambda}}_{\beta} = -{\cal L}_{c}u_\beta$.
The fluid rest-frame cooling rate per unit volume ${\cal L}_{c}$ is determined via
the prescription of \cite{Noble12} in which the gas is cooled at a rate
\begin{equation}
  \mathcal{L}_{c} = \frac{\rho \epsilon}{t_{\mathrm cool}} \left( \frac{\Delta S}{S_0} + \left| \frac{\Delta S}{S_0}\right| \right)^{1/2} \ , 
\end{equation}
where $\Delta S \equiv S - S_0$ and $S = p / \rho^{\Gamma}$ is the local
entropy. Our target entropy, $S_0 = 0.01$, is the initial
entropy of each accretion disk.
Finally, we close the system using a gamma-law equation of state of adiabatic index $\Gamma=5/3$.

\subsection{Grid and Boundary Conditions}
\begin{deluxetable}{l l }
  \tablewidth{\columnwidth}
\tablecolumns{2}
\tablecaption{Warped Grid Parameters \label{tab:grid}}
\tablehead{
  \colhead{Parameter} & \colhead{Value}
}
\startdata
$\delta_{x1}$, $\delta_{x2}$, $\delta_{x3}$ & 0.2 \\
$\delta_{x4}$ & 0.1\\
$\delta_{z}$ & 0.4 \\
$\delta_{y3}$, $\delta_{y4}$ & 0.15 \\
$h_{x1}$, $h_{x2}$, $h_{x3}$, $h_{x4}$, $h_{z}$ & 20.\\
$h_{y3}$, $h_{y4}$ & 10.\\
$s_1$, $s_2$, $s_3$ & 0.01 \\
$b_1$, $b_2$, $b_3$ & 15.\\
$a_{x1}$, $a_{x2}$ & 4.0 \\
$a_{z}$ & 4.3\\
$R_{\mathrm out}$ & $260M$ 
\enddata
\tablecomments{Parameters of the warped grid used for our simulation.
  The full expressions relating these parameters are in Equations 29-32
  of \cite{WARPED}.}
\end{deluxetable}

\begin{figure*}[thb]
  \includegraphics[width=\columnwidth]{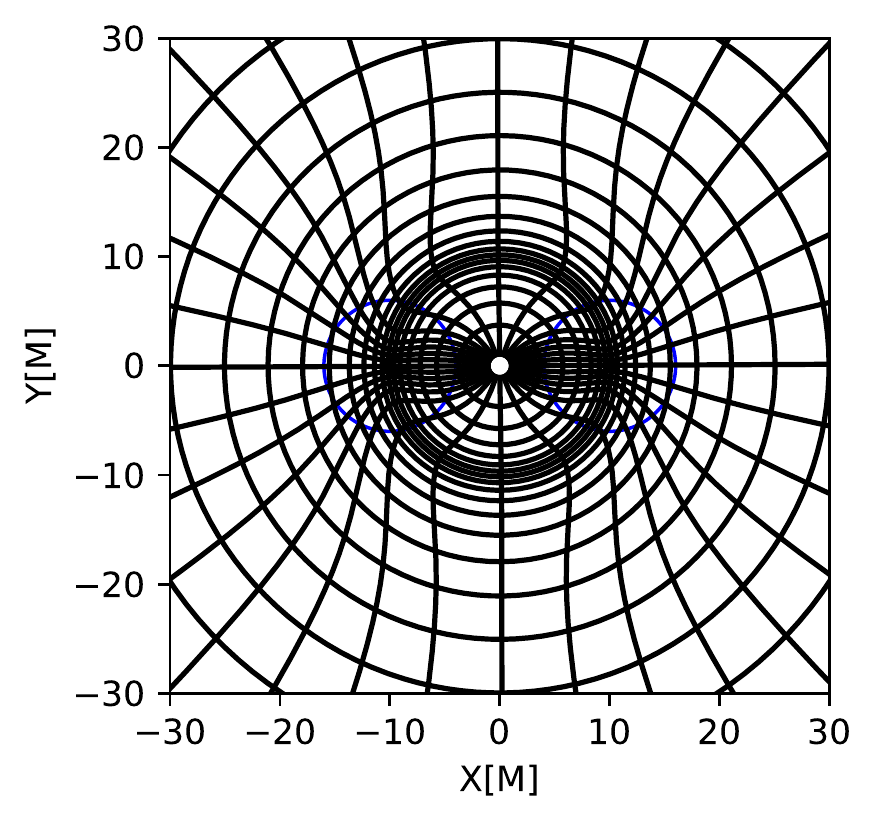}
  \includegraphics[width=\columnwidth,height=0.94\columnwidth]{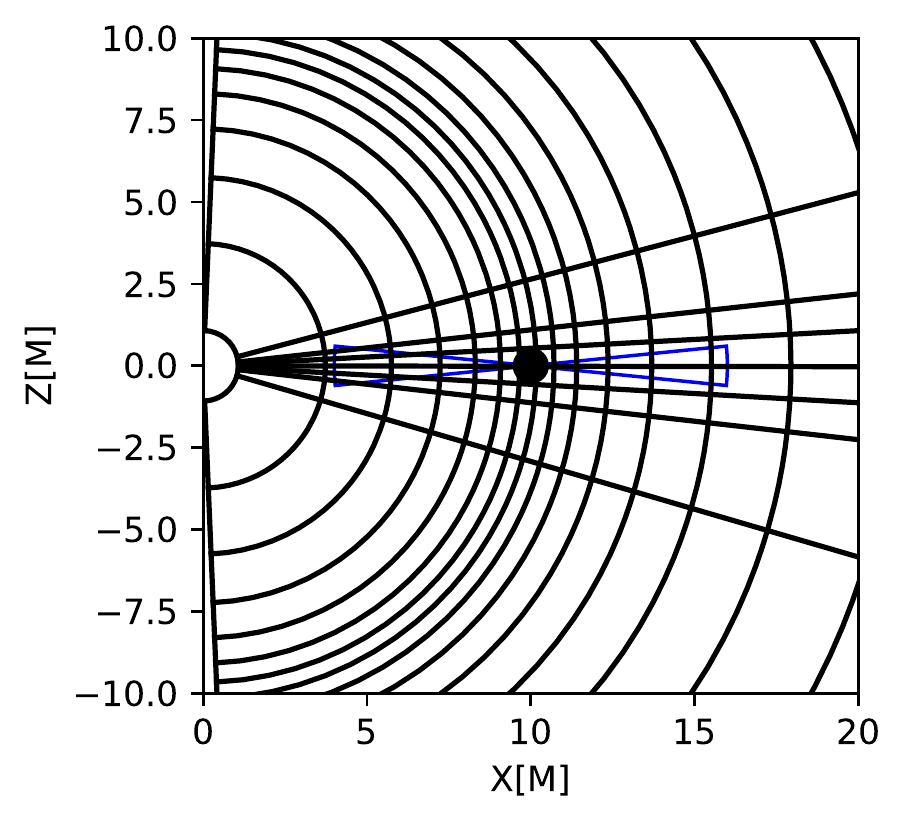}
  \caption{(Left) Equatorial slice of the warped grid used for our simulation at $t=0$.  
    The blue circles denote the Newtonian tidal truncation
    estimate for the mini-disks of $0.3a_0$, where $a_0$ is the initial binary separation. 
    (Right) Poloidal slice of the initial warped
    grid for our simulation. The blue wedges denote the location of a mini-disk of
    constant aspect ratio $\left(H/r\right) = 0.1$ out to the Newtonian tidal truncation estimate. 
    We plot every twentieth grid line. The initial binary separation is $20M$.}
  \label{fig:warped_grid}
\end{figure*}

The simulation performed in \cite{Bowen18}, which we continue here,
is performed in a time-dependent, double fish-eye (warped)
spherical coordinate system whose origin is at the center-of-mass \citep{WARPED}.
We tabulate the full set of grid parameters in Table~\ref{tab:grid}.
This warped gridding scheme facilitates a focusing of cells in the
immediate vicinity of the BHs while preserving spherical coordinates in the circumbinary disk.
The focusing tracks the BHs along their shrinking orbits.
We plot an equatorial and poloidal slice of the grid used in our simulation in Figure~\ref{fig:warped_grid}.
For the poloidal grid, we employ the same polynomial focusing described in~\cite{Noble12}.

We excise the coordinate singularity at the origin by placing a spherical cutout of radius $r=2M$
at the center-of-mass. The cutout size was chosen to sufficiently enlarge the timestep of our
simulation to allow it to be feasible with available computational resources.  Although this cutout
removes the ability to study any transfer of material between the mini-disks, or sloshing~\citep{Bowen17},
it should not impact the physics of the mini-disk--stream--lump interaction presented here significantly.
This is because at any given moment $\lesssim 1\%$ of the combined mini-disk mass is in the
cutout/sloshing region, and in a dynamical time only a fraction of it enters the cut-out~\citep{Bowen17}.
The radial extent of the computational domain was selected as $13a(t=0)$ to fully encompass the circumbinary
disk of~\cite{Noble12} used as initial data. We employ outflow boundary conditions on the radial
boundaries, enforcing that $u^r$ be oriented out of the domain. If not, then we reset the radial
velocity to zero and solve for the remaining velocity components. Our poloidal ``$x^2$'' coordinates have
reflective, axisymmetric boundary conditions at the polar axis cutout and our azimuthal ``$x^3$'' boundary conditions are periodic.

The grid contains $600{\times}160{\times}640$ ($x^1{\times}x^2{\times}x^3$) cells.  In the region of
the circumbinary disk, it exactly matches the grid used in the simulations of~\cite{Noble12}.
Due to the polar grid spacing and off-grid-center location of the BHs, our grid does not include a full
32 cells per scale height in the mini-disks on the side of the BH farthest from the center-of-mass.
Figure \ref{fig:warped_grid} illustrates this asymmetry. However, its cell density is not far below
this number and is sufficient to fully resolve the magnetorotational instability in the circumbinary,
so that MHD effects facilitate the accretion of material into the
central cavity~\citep{Noble12}.
Increasing the poloidal cell count would have further decreased our timestep and required an increase
in the central radial cutout size to compensate. Finally, the azimuthal cell count in the equatorial plane was
selected to be sufficiently large to be comparable to previous hydrodynamic studies~\citep{Bowen17}.


\section{Results}
\label{sec:results}
\subsection{Overview}
\begin{figure}[htb]
  \includegraphics[width=\columnwidth]{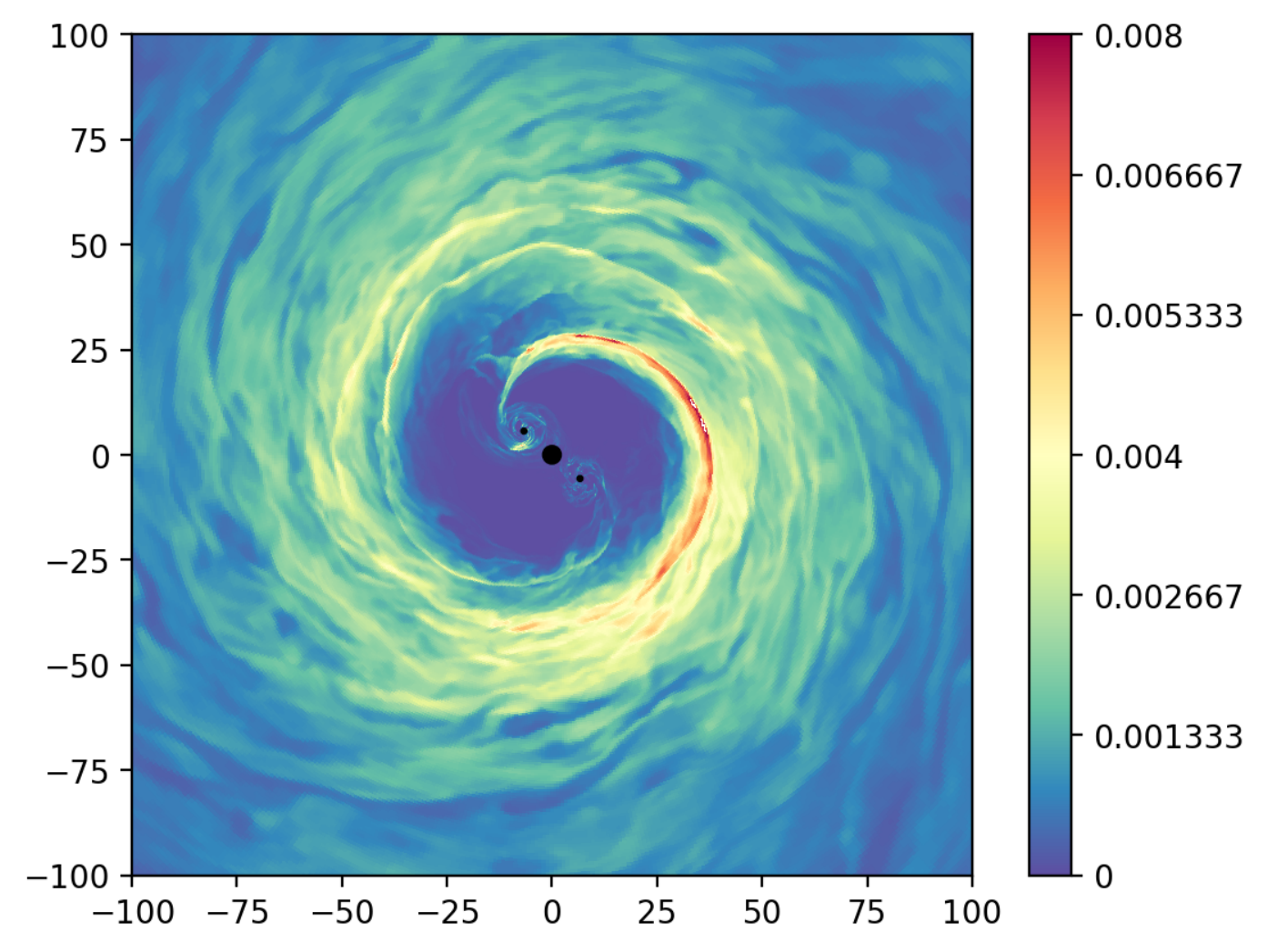}
  \caption{Linear scale density for the innermost $10a_0 \times 10a_0$ at $t=5960M$. The BHs are denoted by
  small black circles. The central cutout is marked by a larger black circle at the center-of-mass.
  Note the one-armed spiral form of the accretion stream.}
  \label{fig:rho_contour}
\end{figure}

Our simulation starts at a binary separation of $20M$, where PN corrections
and inspiral are significant~\citep{2015PhRvD..91b4034Z}. Over the course of the 12 binary orbital periods
simulated here, the binary orbital frequency increases by $\approx 24\%$ and the binary
separation decays by $\approx 15\%$ to $\approx 17M$. 
The lump in the circumbinary disk quasi-periodically modulates the accretion into the
central cavity at twice the BH-lump beat frequency, as each BH comes into orbital phase
with the lump and drains mass from it:
$2\Omega_{\rm beat} = 2\left(\Omega_{\rm bin} - \bar{\Omega}_{\rm lump}\right) \approx 1.44 \Omega_{\rm bin}$
\citep{Shi12,Noble12,DOrazio13,Farris14,Farris15,Farris15a,DOrazio16,Bowen18}.
The association of accretion with the lump creates an asymmetry in the mass flux
carried by the accretion streams, despite the nominal mirror symmetry of the equal-mass binary
(see Figure~\ref{fig:rho_contour})~\citep{Bowen18}.

As was the case in the beginning of the simulation, 
we observe that at any given time throughout the 12 orbits one mini-disk is the beneficiary
of the primary accretion stream, while the other mini-disk is starved of mass supply.
Unlike previous Newtonian studies of mini-disks~\citep{Farris14,Farris15a,DOrazio16},
our binary separation is so small that the tidal truncation radii of the mini-disks are only a factor
of a few times the ISCO radius ($r_{\rm t} \lesssim 2.4 r_{\rm ISCO}$). The radially compact nature
of the mini-disks in our simulation reduces the inflow times to values of order the
BH-lump coupling period~\citep{BI2001,KHH05,Bowen18}. Therefore, during this coupling period one mini-disk is
capable of depleting nearly all of its mass, while the other mini-disk is reformed from significant mass
supply as the accreting stream taps into the over-density of the lump.

Beyond the previously presented three binary orbital periods,
we find an increased clarity of the effects of the dynamic
coupling between the lump and mini-disks. Furthermore, we observe that the state
of the mini-disks throughout the inspiral more closely resembles the most extreme case
presented in ~\cite{Bowen18}, where nearly $75\%$ of the total mini-disk mass can be contained
within a single disk at the peak of the cycle. In Section~\ref{sec:mass-fraction} we present the
long term behavior of the mini-disk refilling and depletion cycle.
To understand this cycle in the context of the mass supply asymmetry, we first study the azimuthal Fourier
structure of the circumbinary disk in Section~\ref{sec:circumbinary_modes}.
Finally, we find that the modulation frequencies of mini-disk mass and the lump are
all tied together through $\Omega_{\rm beat}$ (see Section~\ref{sec:frequencies}).

\subsection{Circumbinary Disk Azimuthal Structure}
\label{sec:circumbinary_modes}
Azimuthal structure can be characterized in terms of its Fourier modes: $f\left(\phi\right) = \sum_m D_m e^{im\phi}$.
We compute the amplitude of the $m$-th component of the circumbinary disk's density as (see e.g.~\citep{Zurek1986,Heemskerk1992})
\begin{equation}
  D_m = \int_{r_{\rm min}}^{r_{\rm max}} \rho \sqrt{-g} e^{-im\phi}d^3x.
  \label{eq:expansion}
\end{equation}
We set our radial inner boundary of the integration as the location of the inner edge of the circumbinary disk,
$r_{\rm min} = 2a(t)$. At large radii, turbulence acts to disrupt
the spiral density structure present within the circumbinary disk. This effectively removes amplitude in $m \neq 0$
while the larger integration volume increases the amplitude of $m=0$. As we are interested in the normalized amplitudes, $D_m / D_0$, we elect to impose
an outer radial boundary on our integration to eliminate $m=0$ noise. The outer radial boundary on our integration is set to $r_{\rm max} = 4a(t)$. 
This is roughly
where the density in the circumbinary disk begins to fall off by eye in our density contours (see Figure~\ref{fig:rho_contour}
at $r \approx 80M$), and is well beyond the $2.4a(t)$ where we expect the lump $m=1$ contribution.

\begin{figure}[htb]
  \includegraphics[width=\columnwidth]{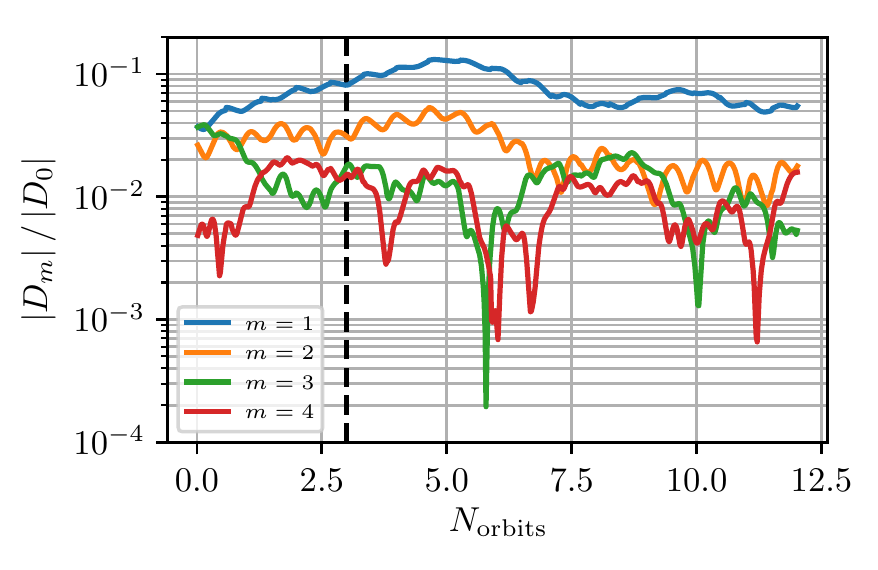}
  \caption{Normalized azimuthal Fourier amplitudes of the circumbinary gas density for $2a(t) \leq r \leq 4a(t)$ from
    the center-of-mass. The vertical line denotes where the previously reported simulation of~\cite{Bowen18} ends.}
  \label{fig:circumbinary_modes}
\end{figure}
As shown in Figure~\ref{fig:circumbinary_modes}, the spectrum of these modes is dominated by $m=1$,
the primary contributor to the lump. Meanwhile, the gravitational quadrupole drives $m=2$ spiral density
waves in the circumbinary disk. The normalized strength of the $m=1$ component can be as much as five
times greater than the $m=2$ component at the same time. Next, although it is beyond the scope of this
paper, we note that there appear to be high frequency oscillations within the dominant $m=1$ and $m=2$
components. Finally, there may also be a low frequency modulation in the overall strength of the lump itself,
much like that seen in the simulation of \citet{Shi12}.

\begin{figure}[htb]
  \includegraphics[width=\columnwidth]{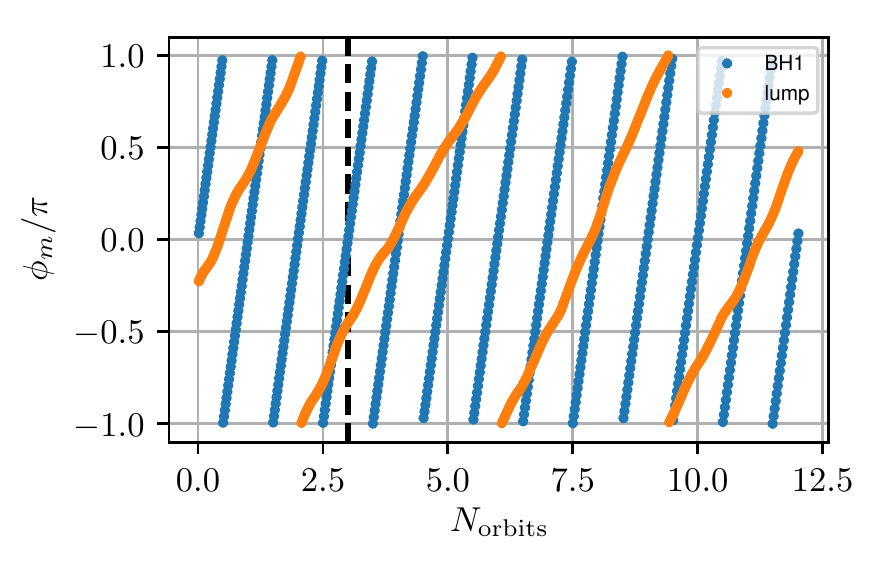}
  \includegraphics[width=\columnwidth]{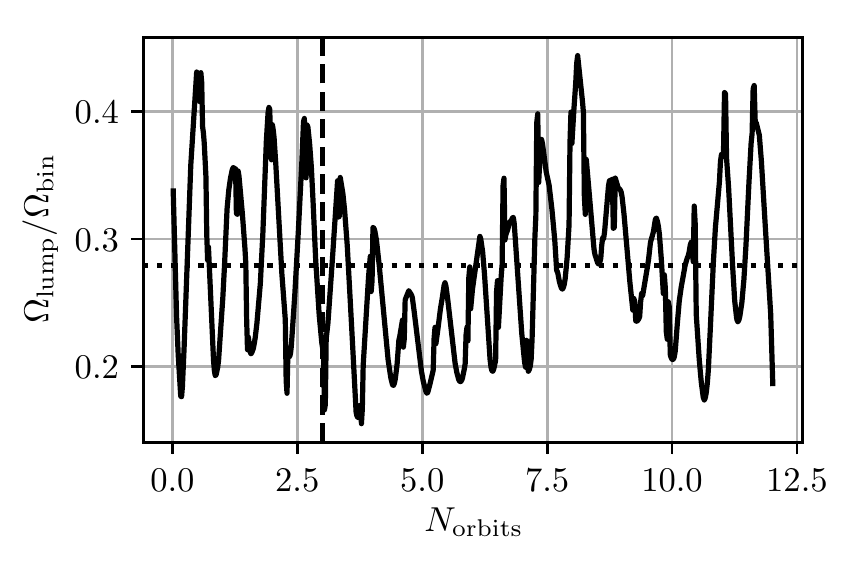}
  \caption{(Top) BH and lump phases throughout the simulation. (Bottom) The $m=1$ pattern speed,
    its instantaneous angular velocity, normalized to 
    the binary orbital frequency as a function of time in our simulation.
    The horizontal dotted line denotes the average value of 0.28. The vertical line denotes where the previously reported simulation of~\cite{Bowen18} ends.}
  \label{fig:pattern_speed}
\end{figure}

The primary advantage of our Fourier series expansion is that we can now use it to track the phase of the lump in time. We calculate
the lump phase with the $m=1$ azimuthal pattern location as \citep{Zurek1986,Heemskerk1992}
\begin{equation}
  \phi_{m} = \tan^{-1}\left(-\frac{\Im\left(D_m\right)}{\Re\left(D_m\right)}\right),
\end{equation}
where $\Im\left(D_m\right)$ and $\Re\left(D_m\right)$ are the imaginary and real parts of the $m$ component of our Fourier series expansion.
In the top frame of Figure~\ref{fig:pattern_speed} we plot $\phi_m$ for the lump and the phase of one of the BHs as a
function of time for our simulation. A point where the two lines cross would correspond to a BH coming into
phase with the lump.

In Figure~\ref{fig:pattern_speed} we observe that the lump phase is not advancing at a constant rate.
To highlight this, we define the instantaneous lump angular velocity as the $m=1$ pattern speed,
\begin{equation}
  \Omega_{\rm lump} = \partial_t \phi_{m=1} .
\end{equation}
We plot this angular velocity in ratio to the instantaneous binary orbital frequency in the bottom
frame of Figure~\ref{fig:pattern_speed}. First, we note that
there does not appear to be any significant secular trend in the ratio $\Omega_{\rm lump} / \Omega_{\rm bin}$. This
implies that the lump is readily able to track the binary inspiral throughout our simulation.
Second, there are quasi-periodic oscillations in the ratio that
perturb it by as much as 30--40\% from its mean value $\approx 0.28$.
Finally, we observe that the mean value of the ratio between the lump's angular velocity and the binary
orbital frequency is very close to the ratio that would obtain if the lump followed a Keplerian orbit
with semi-major axis $2.4a(t)$, i.e., $0.27$.  On this basis, as well as seeing that the lump's density concentration
moves very coherently, we argue that the lump is better thought of as a physical object following an orbit
rather than a pattern through which orbiting fluid moves.

\subsection{Mini-Disk Refilling and Depletion Cycle}
\label{sec:mass-fraction}
\begin{figure}[htb]
  \begin{center}
    \includegraphics[width=0.9\columnwidth]{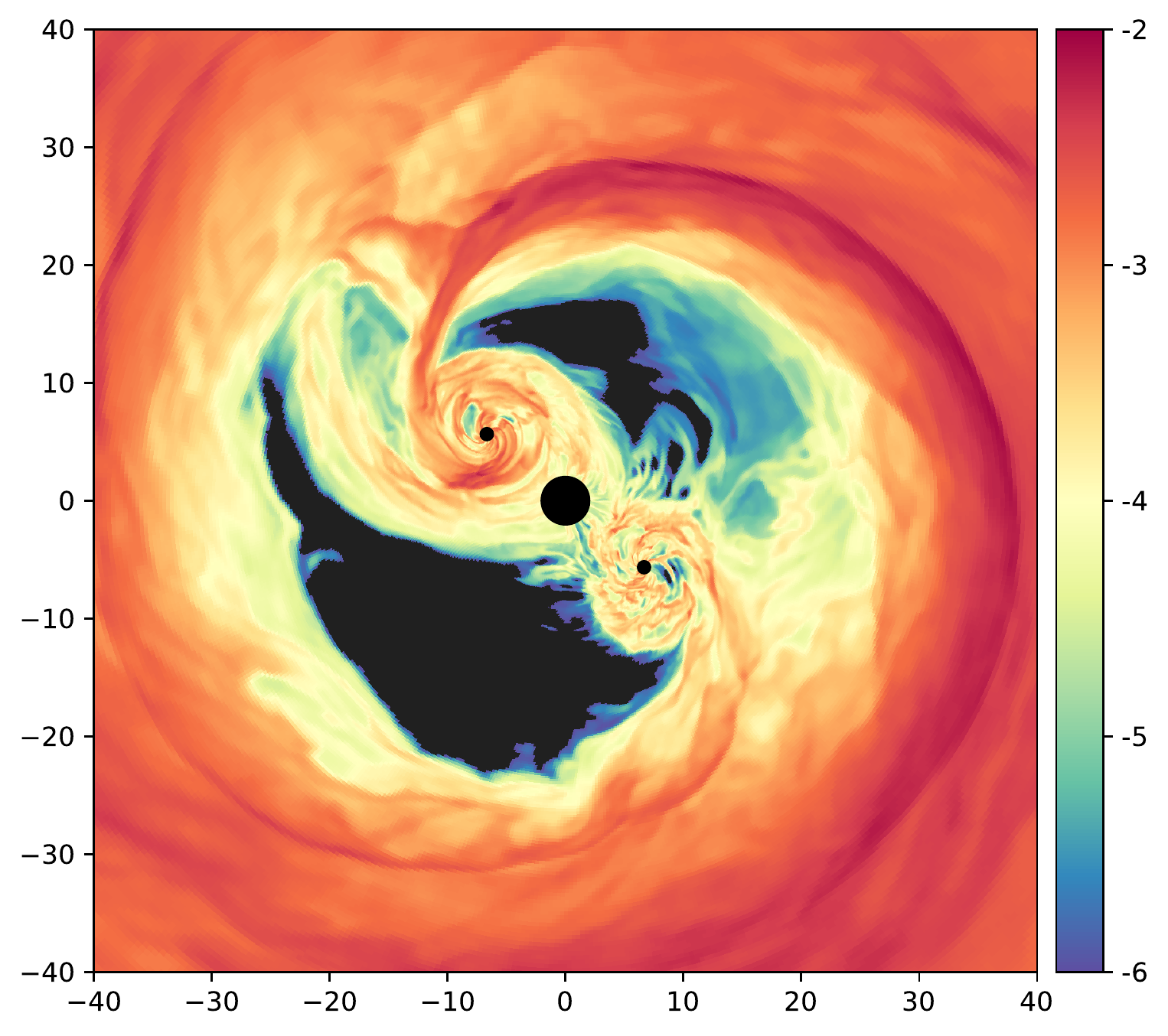}
  \end{center}
  \caption{Logarithmic scale density for the innermost $4a_0 \times 4a_0$ at $t=5960M$. The BHs are denoted by
  small black circles. The central cutout is marked by a larger black circle at the center-of-mass.}
  \label{fig:rho_contour_log}
\end{figure}
We plot an equatorial slice of the gas density at $t=5960M$ in
Figures~\ref{fig:rho_contour}~and~\ref{fig:rho_contour_log}. The latter figure shows clearly how
large the contrast in mass between the two disks can be, a contrast stemming from the alternation in feeding
between the two.  In Figure~\ref{fig:rho_contour}, a dense band can be seen coming off the inner edge of
the circumbinary disk on the right side and connecting to the BH on the left.  This mini-disk has just past
its time of maximum mass, so its mass supply, having exceeded its internal accretion rate for the past
$\approx (\pi/2)/\Omega_{\rm beat}$ has now fallen a little below balance with its internal accretion rate.
Conversely, the left side of the circumbinary edge produces a significantly less dense stream that deposits material
onto the BH on the right. At the moment illustrated in the figures, the mass supply for this BH is
beginning to overcome accretion mass-loss, so the mass of this disk is starting to grow.

\begin{figure}[htb]
  \includegraphics[width=\columnwidth]{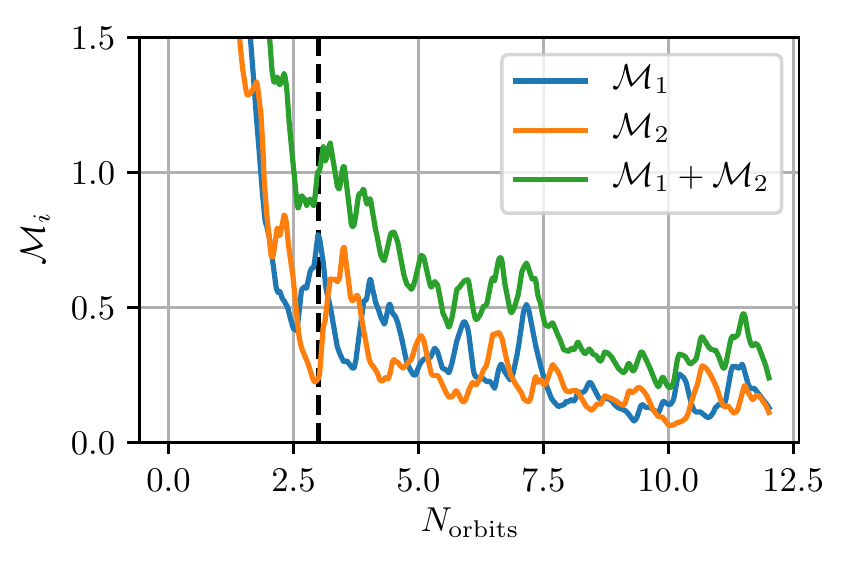}
  \includegraphics[width=\columnwidth]{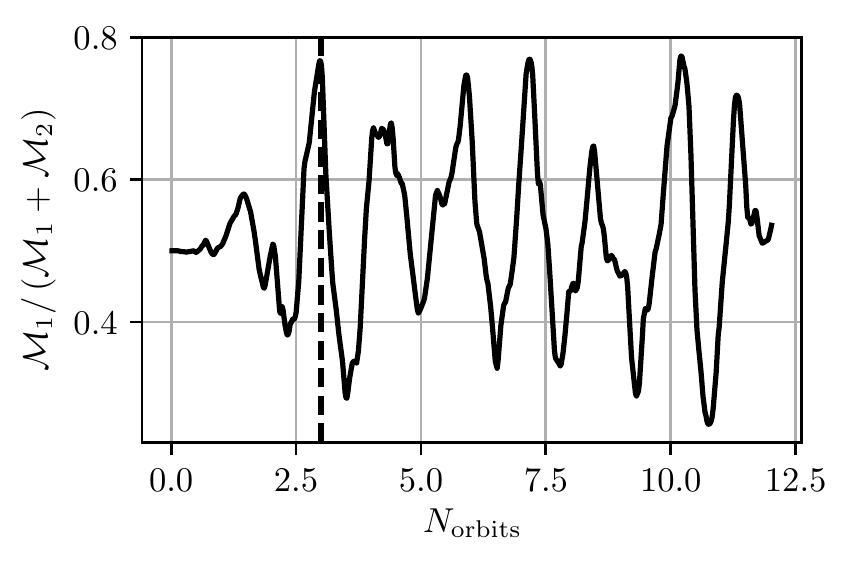}
  \caption{(Top) Total integrated mass in each mini-disk and combined total mini-disk mass.
    (Bottom) Fraction of total mini-disk mass contained within mini-disk one.
    Time is in units of binary orbital periods.
    The vertical line denotes the end of the simulation previously reported in~\cite{Bowen18}.}
  \label{fig:mini-disk-masses}
\end{figure}

The result of the asymmetric inflow into the central cavity is that the mini-disks exist in a persistent
state of inflow disequilibrium, with their masses constantly increasing and decreasing. This can be most
readily observed by looking at the fraction of total mini-disk mass that exists within a single mini-disk
at a time. We define the mass contained
within a mini-disk ($\mathcal{M}_i$) from the continuity equation as
\begin{equation}
  \mathcal{M}_i = \int_{r=m_i}^{0.4a(t)} \rho u^0 \sqrt{-g} d^3x,
\end{equation}
where $r=m_i$ is the radial location of the horizon for our non-spinning BH of mass $m_i$ in the PN Harmonic coordinates
which we evolve in. We integrate over the full solid angle in angular coordinates.
In Figure~\ref{fig:mini-disk-masses}, we plot each mini-disk mass, the combined mini-disk mass,
and $\mathcal{M}_1 / \left(\mathcal{M}_1 + \mathcal{M}_2\right)$
as a function of time for our simulation.

Observing the top panel of Figure~\ref{fig:mini-disk-masses}, we note that the overall secular decay in
the mini-disk masses appears to level out by the end of our simulation. Next, we note that there are
quasi-periodic modulations in each mini-disk mass and the total mini-disk mass. However, neither
of the oscillation periods of the combined mini-disk mass matches the oscillation period of
the individual mini-disk masses (see
Section~\ref{sec:frequencies}). Finally, we  observe that while the mass contained within one mini-disk
increases, the mass contained within the other mini-disk decreases. This leads to quasi-periodic fluctuations
in the fractional mass contained within a single mini-disk.

We now turn our attention to the mini-disk mass fractions in the bottom panel of Figure~\ref{fig:mini-disk-masses}. 
First, at least $70\%$ of
the total mini-disk mass is contained within a single mini-disk at six times throughout the simulation.
Second, the quasi-periodic nature of the mini-disk filling/depletion cycle is fairly regular throughout the simulation and significantly more
noticeable than in the first few orbits. Finally, the degree to which the mass fraction oscillates through our simulation does not appear
to have significant dependence on binary separation.

\subsection{Frequencies of Quasi-Periodic Behavior}
\label{sec:frequencies}
\begin{figure}[htb]
  \includegraphics[width=\columnwidth]{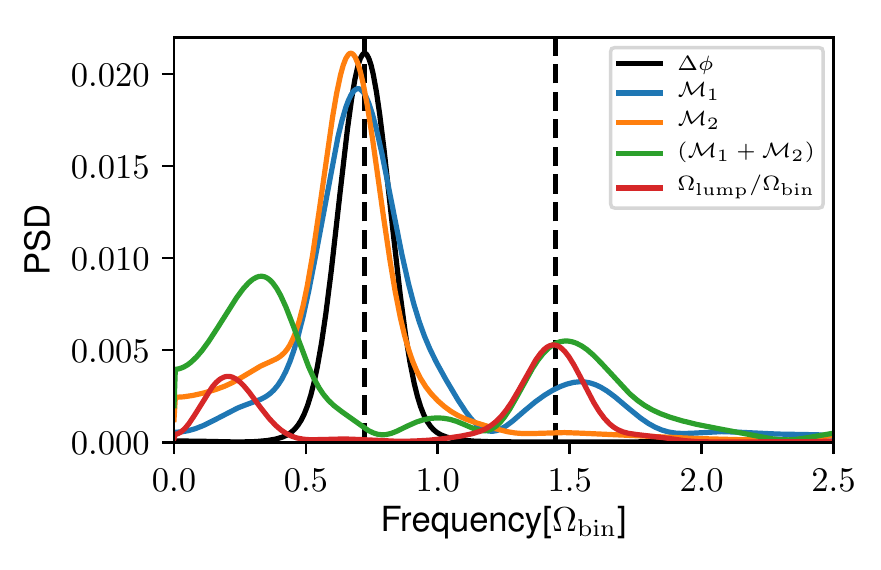}
  \caption{Power Spectral Density of various quantities in our simulation. We use a hamming window
    and Welch PSD algorithm. A linear fit
    is subtracted from the raw data before calculating the PSD. 
    Frequency is in units of average binary orbital frequency.
    The beat frequency is calculated using $\Delta \phi = \left| \phi_{\rm lump} - \phi_{\rm BH} \right|$ for all time.
    All other quantities are calculated from $N_{\rm orbits} = 2$ to the end of the simulation to remove
    noise associated with secular equilibration. The vertical dash lines denote $\Omega_{\rm beat}$
    and $2\Omega_{\rm beat}$.}
  \label{fig:psd}
\end{figure}

Just as the azimuthal Fourier modes cleanly revealed azimuthal spatial structure, the power spectral
density (PSD) picks out the principal timescales of variation.  From Figure~\ref{fig:psd}, we see
that the time-dependence
of four properties of the system---total mass of the mini-disks, the two individual mini-disk masses,
and the orbital frequency of the lump---can all be described in terms of periodic behavior at three
frequencies: the orbital frequency of the lump, the beat between the lump's orbital frequency and the
binary orbital frequency, and twice the beat frequency.  In other words, the interaction of the lump
with the binary orbit drives the mass of the mini-disks.

The total mini-disk mass is modulated at a frequency slightly greater than the lump orbital frequency
and at twice the beat frequency, but not at $\Omega_{\rm beat}$, the oscillation frequency of the
individual mini-disks.  The reason is that the symmetry of the binary makes the filling/depletion
cycles for the two BHs exactly $\pi$ out of phase and of roughly equal amplitude.
However, the total mass supply into the central cavity is tied to phase-matching the lump and either
partner in the binary.  The total mass of the mini-disks can oscillate because the lump's injection
of mass each time it comes into phase with a BH is irregular, and the inflow time in each
mini-disk is so short ($\sim 2\pi/\Omega_{\rm beat}$) that inflow equilibrium within
the mini-disks is never achieved.

The two mini-disk masses vary primarily at the beat frequency, $\approx 0.72 \bar{\Omega}_{\rm bin}$,
where $\bar{\Omega}_{\rm bin}$ is the average binary orbital frequency from $N_{\rm orbits}=2$ to the
end of the simulation (the same range over which the PSD is calculated).  As already pointed out,
this value corresponds to the mean beat frequency
$\Omega_{\rm beat} = \bar{\Omega}_{\rm bin} - \bar{\Omega}_{\rm lump}$, when $\bar{\Omega}_{\rm lump}$
is taken from Figure~\ref{fig:pattern_speed}.  Although there is power outside of this peak in the PSD
for $\mathcal{M}_1$, we do not believe this to be a robust signal because the location and
structure of this peak depend on the starting point of data we include in the PSD calculation.

The lump orbital frequency (in units of the binary orbital frequency) is itself modulated at two frequencies.
The low frequency modulation in the lump motion appears at $\approx 0.2\bar{\Omega}_{\rm bin}$. Due to
the close proximity to the orbital frequency of the lump, we speculate that this modulation may be
a result of eccentricity of the lump's orbit in the circumbinary disk~\citep{Shi12}.

The lump motion's high frequency modulations correspond to precisely twice the beat frequency, or
the frequency at which the lump comes into phase with either BH. We speculate that this modulation
is due to the time-dependent gravitational force exerted by a BH onto the lump. As an individual BH
starts coming into phase with the lump, it is gravitationally pulled toward the BH in the opposite direction
of the lump orbital velocity. Since the BH orbits more quickly than the lump, it then passes the
lump and exerts a gravitational pull in the same direction of the lump orbital velocity. This could
result in a gravitationally-driven slow down and then reacceleration of the lump with each BH passing
at twice the beat frequency, $\approx 1.44 \Omega_{\rm bin}$. Alternatively, this could be the result
of shocks when the streams are flung back into the lump (which would also occur at twice the beat frequency).

\section{Discussion}
\label{sec:discussion}

\subsection{Quasi-Periodic Oscillations in Relativistic Binaries}
\label{sec:psd-discussion}
Our simulation finds that the dynamics of the central cavity and circumbinary
are tangled together, with one informing the evolution of the other
in a feedback cycle. As the binary performs many orbits, the azimuthal structure
of the circumbinary disk becomes increasingly more $m=1$
dominant~\citep{Noble12,Shi12,DOrazio13,Farris14,Farris15,Farris15a,DOrazio16}.
However, previous simulational studies \citep{Farris14,Farris15,Farris15a,DOrazio16,2018MNRAS.476.2249T}
have all reported that the individual mini-disks around each BH should be nearly symmetric for $q=1$ binaries. 

Our results contrast with the previous work because our binary has such a small separation;
there is then only a short distance between the outer and inner radii of the mini-disks
($r_{\rm t} \lesssim 2.4 r_{\rm ISCO}$), and their inflow times become as short as the binary period.
As a result, the quasi-periodic
modulations in the mass supply into the central cavity directly imprint themselves onto
the total mass within the central cavity, and therefore the rest-mass energy available for photon
radiation. For example, the primary modulations in
mini-disk mass occurs at the frequency of their respective BHs coming into phase with the lump.
Meanwhile, the total combined cavity mass is modulated at the same frequencies as the lump orbital
frequency.
This seems to imply that the binary evolution 
significantly impacts the mini-disk evolution via binary--lump and lump--mini-disk coupling.

Remarkably, after a transient period lasting $\sim 3$ orbits, our simulation settles into a very
regular behavior pattern in which all characteristic timescales maintain constant ratios even while
the fundamental timescale, the binary orbital period, evolves.  The longest timescale, the orbital
period of the lump, is $\propto a^{3/2}$ exactly like the
binary orbital period: this determines the slow modulation of the lump's angular velocity and
the slow modulation of the total mini-disk mass.  The middle timescale is the inverse of the beat frequency,
and it governs the modulation of the individual mini-disk masses.
The beat frequency is $\Omega_{\rm bin} - \bar{\Omega}_{\rm lump}$; with both pieces $\propto a^{-3/2}$,
the inverse is also $\propto a^{3/2}$.   The shortest timescale, accounting for the rapid modulation of
both $\Omega_{\rm lump}$ and the total mass, is half the inverse beat frequency, so it is likewise $\propto a^{3/2}$.
Even the inflow time is related to $a^{3/2}$ because the tidal truncation radius $r_t \propto a$ in both Newtonian
\citep{Paczynski:1977,Papaloizou:1977a,LinPap:1979,ArtymLubow94,Mayama10,VB11,Nelson16} and relativistic
conditions~\citep{Bowen17}, and the inflow time is most closely tied to the orbital period at the
mini-disk's outer edge.

However, if the rapid acceleration in orbital evolution due to gravitational radiation causes the binary
to shrink faster than the inner edge of the circumbinary disk, the ratio
$\left(\Omega_{\rm lump}/\Omega_{\rm bin}\right)$ would exhibit a secular trend toward smaller values.
This would increase $\left(\Omega_{\rm beat}/\Omega_{\rm bin}\right)$, and
therefore drive more rapid cycles of mass supply and deprivation in the mini-disks.

During this evolution, the inflow rate in the mini-disks also accelerates because it is $\propto a^{-3/2}$.
We expect, therefore, that the mini-disk masses would continue to follow the cycle of filling and depleting
much as they have during our simulation.
Eventually, however, tidal truncation of the mini-disks will come at such a small radius that
they can no longer exist: $r_{t} = r_{ISCO}$ when $a \approx 8.3M$ for non-spinning BHs

As a final comment, we note that the precise frequencies, while relatively constant in units of $\Omega_{\rm bin}$,
vary significantly throughout our simulation as a result of inspiral. Over the span of 12 binary orbital periods,
the binary orbital frequency increases by as much as $\approx 24\%$ of its initial value. In addition to
an increased binary orbital frequency, the binary separation (and therefore tidal truncation radii) have shrunk
by $15\%$ of their original values.
These significant changes in binary orbital parameters account for the broadened
peak structure exhibited in Figure~\ref{fig:psd}.
Furthermore, our PSD results could be weighted more heavily towards the values at larger separations. This is
because the binary spends significantly more time at larger separations than smaller due to the non-linear
growth in inspiral rate from gravitational waves.  It is possible that electromagnetic counterparts
for merging SMBBHs, particularly in the high-energy bands emanating from the mini-disks, could be marked by
a ``chirp-like'' growth in quasi-periodic fluctuations \citep{2017PhRvD..96b3004H}.

\subsection{Possible Parameter Dependence}
\label{sec:parameters}
The magnitude of mini-disk oscillations reported here depends on a number of
parameters.  The mini-disk cycle depends on binary separation,
binary mass ratio, BH spin, and the accretion rate, which determines the run of
temperature across the mini-disks and the circumbinary disk. It would therefore be presumptuous to claim
that the oscillations reported here are completely generic to relativistic binaries.

Although the lump appears to be a generic result of circumbinary accretion simulations
~\citep{Noble12,Shi12,DOrazio13,Farris14,Farris15,Farris15a,DOrazio16,2018MNRAS.476.2249T},
its quantitative structure remains subject to some uncertainty.  In particular,
3D MHD simulations of circumbinary disks around non-evolving binaries have
shown secular growth in its amplitude over timescales $\sim 10^2$ orbits \citep{Shi12,Noble12}.
The amplitude of the lump in our simulation, drawn from a somewhat arbitrary moment
in the simulation of \citet{Noble12}, may not necessarily match the amplitude of
the lump found in a binary with separation $a=20M$ that evolved over a much longer
period of time from an initial state of much greater separation.  On the other hand, pseudo-Newtonian
2D hydro simulations with a phenomenological viscosity can follow circumbinary disks
for much longer times: for example, all the way from $30M$ separation to merger \citep{2018MNRAS.476.2249T},
albeit with severe physics limitations.  The fact that the
lump amplitude they find (factor of several contrast in surface density) is
similar to what we find suggests that the amplitude may be relatively insensitive
to the arbitrary assumptions of different simulations.

The binary mass ratio could also play a significant role in the results presented here.
Previous studies have shown that the strength of the lump is a function of the mass ratio of
the binary. As the mass ratio deviates further from unity, the overall size of the lump
in 2D hydro simulations diminishes~\citep{Farris14,DOrazio16}.
This effect continues until a binary mass ratio of $q = 0.04$, where the structure of the
central cavity is altered dramatically~\citep{DOrazio16}.
More recent studies including MHD stresses find that the lump is only significant
  for mass ratios $q \geq 0.2$ ~\citep{LUMP}. 

In addition to altering the strength of the lump and mass supply fluctuations, the binary
mass ratio will lead to preferential accretion onto the secondary~\citep{Farris14,Shi2015}. In this scenario,
the quasi-periodic oscillations in each mini-disk mass would not be the result of a flipping back
and forth between each mini-disk coupling to the lump presented here, but rather a quasi-periodic
modulation of mass supply to the secondary. In other words, the modulations in mass of
the mini-disk around the more massive binary member would likely be diminished.
Conversely, the modulations in the 
mini-disk around the secondary could exhibit even more pronounced fluctuations. In addition,
the binary mass ratio would alter the tidal truncation radii of each mini-disk (potentially
creating larger ratios of $r_{\rm t}/r_{\rm ISCO}$ for the primary's
mini-disk and smaller fractions for the secondary's mini-disk).

There is, however, a possible self-regulation of the mass-ratio.  Preferential accretion onto
the secondary directly drives the mass ratio toward unity.  In the relativistic regime, it is
possible that there may be net transfer due to sloshing that acts in the same direction.

Another important caveat is that astrophysical BHs are expected to have spin~\citep{1969Natur.223..690L}.
The dominant effect of relevance is the modification to $r_{\rm ISCO}$~\citep{1972ApJ...178..347B}.
If the angular momentum axis of the circumbinary
disk, and therefore the angular momentum axis of the mini-disk, is aligned with the BH spin axis,
the ratio of $r_{\rm t}/r_{\rm ISCO}$ will increase~\citep{Campanelli:2006uy}. This would serve to
increase the mini-disk inflow times and prevent the mini-disk from fully depleting while starved
of mass supply. It is therefore conceivable that spin effects could delay the onset of the mini-disk
cycle reported here to smaller binary separations.  Conversely, if the mini-disk
angular momentum axis were counter-aligned with the BH spin axis, the ISCO would grow. This would either
decrease the mini-disk inflow times, or prevent the formation of a mini-disk at larger
binary separations.  Finally, with certain misaligned spin configurations, the BH spin axes have
been shown to change direction~\citep{Campanelli:2006fy}, or flip-flop, during the final stages of inspiral
for mass ratios $q \geq 0.5$~\citep{2015PhRvL.114n1101L,2016PhRvD..93l4074L,2016PhRvD..93d4031L,Gerosa2018}. Such 
binary dynamics could have a profound effect on the mini-disk structure.

Furthermore, the internal stresses of the mini-disks and circumbinary disk depend on the disk temperature
and resulting disk thickness. Our mini-disks may be hotter and thicker than mini-disks in nature,
and therefore may have unphysically short inflow periods.
However, astrophysical mini-disks will still be very hot. These high temperatures would serve to augment
secondary accretion processes such as spiral shocks 
\citep{Ju16,Bowen17,RyanMacFadyen17,Bowen18}.

As we have outlined, the effects presented here depend strongly on the ratio $r_{\rm t} / r_{\rm ISCO}$,
and therefore the binary separation. At larger separations, such as those in Newtonian studies,
the inflow periods within the mini-disks will be too long for any significant depletion in
the beat period. The binary separation where this mini-disk cycle takes hold remains unknown.

\section{Conclusions}
\label{sec:conclusions}
In this paper we have extended the first ever study of magnetized mini-disks
coupled to circumbinary accretion in a relativistic SMBBH approaching merger.
By extending the simulation to four times the original number of binary orbital
periods, we were able
to extract PSDs of the quasi-periodicity reported in~\cite{Bowen18} and explore
longer term behavior through inspiral.

We found that nearly every aspect of our late-inspiral SMBBH can be characterized
by quasi-periodic oscillations associated with the interaction of the individual BHs
and the lump in the circumbinary disk. The individual BHs interact with
the circumbinary disk by imparting angular momentum into the streams, flinging
them into the circumbinary disk and forming the lump. The lump then interacts
with the BHs by quasi-periodically modulating the mass accretion into the central
cavity at frequencies of $2\Omega_{\rm beat} \approx 1.44 \Omega_{\rm bin}$ and $\approx 0.2 \Omega_{\rm bin}$.

Because of the small binary separation, the radial extent of our mini-disks is
$\lesssim 2.4 r_{\rm ISCO}$. This leads to rapid depletion timescales in the mini-disks
of order the period associated with an individual BH interacting with the lump: $\Omega_{\rm beat} \approx 0.72 \Omega_{\rm bin}$.
Furthermore, because the individual mini-disk masses oscillate at the beat frequency and deplete nearly
all of their mass over this timescale before recoupling to the lump, we observe that the total
mass within the mini-disks quasi-periodically oscillates at the same frequencies as the lump.

An astrophysical SMBBH resembling our simulation would exhibit quasi-periodicity in response
to the strong fluctuations in the quantity and structure of gas in the immediate vicinity of
the BHs. These quasi-periodic oscillations would
be intimately tied to the structure and dynamics of the lump, and would augment Doppler variability.
Finally, because our mini-disks live far outside the analytic regime of inflow equilibrium, our
results emphasize the need for generating self-consistent light curves using non-linear
MHD simulations.

\section*{Acknowledgments}
D.~B.~B. is supported by the US Department of Energy through the Los Alamos 
National Laboratory. Los Alamos National Laboratory is operated by Triad National 
Security, LLC, for the National Nuclear Security Administration of U.S. 
Department of Energy (Contract No. 89233218CNA000001).
For this work, D.~B.~B. also acknowledges partial support from NSF grants
AST-1028087, AST-1516150, PHY-1305730, PHY-1707946, OAC-1550436 and
OAC-1516125. M. A. is a Fellow of the RIT's Frontier of Gravitational Wave Astronomy.
M. A., M.~C. and V.~M. received support from NSF grants
AST-1028087, AST-1516150, PHY-1305730, PHY-1707946, OAC-1550436 and
OAC-1516125. S.~C.~N. was supported by AST-1028087, AST-1515982 and
OAC-1515969, and by an appointment to the NASA Postdoctoral Program at
the Goddard Space Flight Center administrated by USRA through a
contract with NASA.  J.~H.~K. was partially supported by NSF grants
AST-1516299, PHYS-1707826 and OAC-1811287, as well as Simons Foundation
grant 559794. V.M. also acknowledges partial support from AYA2015-66899-C2-1-P.

Computational resources were provided by the Blue Waters
sustained-petascale computing NSF projects OAC-1811228,
OAC-0832606, OAC-1238993, OAC-1516247 and OAC-1515969, OAC-0725070. 
Blue Waters is a joint effort of the University of Illinois at Urbana-Champaign and its
National Center for Supercomputing Applications. Additional resources
were provided by XSEDE allocation 
TG-PHY060027N and by the BlueSky Cluster at Rochester Institute of Technology.  
The BlueSky cluster was supported by NSF grants AST-1028087, PHY-0722703 and PHY-1229173.

\hfill \break
\bibliography{bhm_references}

\end{document}